\documentclass[fleqn,twoside]{article}
\usepackage{amsmath}
\usepackage{amsfonts}
\usepackage{amssymb}
\usepackage{graphicx}%
\def\be{\begin{equation}}

\def\ee{\end{equation}}
\def\bes{\begin{equation}\begin{split}&}
\def\es{\end{split}}
\def\bi{\bibitem}
\begin{document}

\title{Canonical formulation of Pais-Uhlenbeck action and resolving the issue of branched Hamiltonian.}
\author{Kaushik Sarkar$^\dag$, Nayem Sk$^\ddag$, Ranajit Mandal$^{*}$ and  Abhik Kumar Sanyal$^{\S}$\\
$^\dag, ^\ddag, ^*, $ Dept. of Physics, University of Kalyani, Nadia, India - 741235\\
$^{\S}$Dept. of Physics, Jangipur College, Murshidabad, India - 742213}
\maketitle
\begin{abstract}
Canonical formulation of higher order theory of gravity requires to fix (in addition to the
metric), the scalar curvature, which is acceleration in disguise, at the boundary. On the contrary, for
the same purpose, Ostrogradski’s or Dirac’s technique of constrained analysis, and Horowitz formalism,
tacitly assume velocity (in addition to the co-ordinate) to be fixed at the end points. In the process
when applied to gravity, Gibbons-Hawking-York term disappears. To remove such contradiction and
to set different higher order theories on the same footing, we propose to fix acceleration at the endpoints/
boundary. However, such proposition is not compatible to Ostrogradski’s or Dirac’s technique.
Here, we have modified Horowitz’s technique of using an auxiliary variable, to establish a one-to-one
correspondence between different higher order theories. Although, the resulting Hamiltonian is related
to the others under canonical transformation, we have proved that this is not true in general. We have
also demonstrated how higher order terms can regulate the issue of branched Hamiltonian.
\end{abstract}
\footnotetext[1]{
\noindent Electronic address:\\
$^\dag$sarkarkaushik.rng@gmail.com\\
$^\ddag$nayemsk1981@gmail.com\\
$^*$ranajitmandalphys@gmail.com\\
$^{\S}$sanyal\_ ak@yahoo.com\\}

\section{Introduction}
We often encounter higher order theories, for example, in radiation reaction \cite{1}, noncommutative quantum field theories \cite{2}, anyons (quasi particles appearing in two-dimensional systems, which are neither fermions nor bosons) \cite{3} and particularly in the context of gravitation, leading to field equations with fourth order or even more derivatives. A quantum theory of gravity in any of its form, (eg., arising in the weak energy limit of a yet complete theory, may be the string theory, M-theory or supergravity) contains higher order curvature invariant terms leading to higher order theory of gravity. Higher order theories also appear in classical mechanical problems such as Pais-Uhlenbeck oscillator \cite{Pais}. A host of canonical formalisms appear in the literature to handle higher order theory. Some of these lead to the same phase-space Hamiltonian, some different. Although, all these Hamiltonian lead to correct classical field equations, some of these are very different and not related to the others through canonical transformations (see appendix). Further, there happens to be a mismatch in different canonical formalisms as far as boundary terms are concern. Any higher order theory is associated with boundary (end point) terms which do not vanish fixing only the co-ordinate at the end points. For fourth order theory, it is required to fix either the velocity or the acceleration at the end-points (boundary) in addition. However, it does not mismatch Cauchy data with the endpoint (boundary) data. This is because, canonical formulation of higher order theory requires additional degree of freedom. Ostrogradki's technique \cite{Ostro}, Dirac's constrained analysis \cite{Dirac} and Horowitz's formalism (originally developed for the theory of gravity) \cite{Horo} fix velocity at the end points. \\

In case of gravity, no nontrivial lagrangian $\mathcal{L}_g$ can be constructed from the metric $g_{\mu\nu}$ and its first derivatives alone. Even for Einstein-Hilbert action, Lagrangian depends on second derivatives of the metric, in the form $\mathcal{L}_g(g, \partial g, \partial^2 g)$. Under metric variation, setting $\delta g_{\mu\nu}|_{\partial\mathcal{V}} = 0$, Einstein-Hilbert action yields a surface term, so that the complete action reads

\be\label{A1} A_1 = \int \Big[{R-2\Lambda\over 16\pi G} + \mathfrak{L_m}\Big]\sqrt{-g}d^4 x + {1\over 8\pi G}\oint_{\partial\mathcal{V}} K \sqrt{h}d^3 x,\ee
and the supplementary surface term is known as Gibbons-Hawking-York (GHY) boundary term \cite{Boun}. In the above, $\mathfrak{L_m}, K ~\mathrm{and}~ h$ are the matter Lagrangian density, the trace of the extrinsic curvature tensor and the determinant of the three-space metric respectively. Note that, one is not allowed to fix the extrinsic curvature tensor $(K_{ij})$ at the boundary to get rid of the GHY term. This is because, the correspondence between Cauchy initial value problem of field equations and the boundary value problem arising out of variational principle requires that the number of initial data should coincide with the number of boundary data. If $K_{ij}$ is fixed at the boundary, boundary data exceeds the number of configuration space variables, which mismatch Cauchy data. Canonical formulation of general theory of relativity (GTR) has been performed successfully, taking into account the GHY term, and is known by the name of ADM formalism \cite{ADM}. \\

Canonical formulation of higher order theory of gravity requires additional degrees of freedom viz. ($h_{ij}, K_{ij}$) \cite{Boul}, as already mentioned. Metric variation of higher order theory of gravity in its simplest form, viz., $f(R) \propto \alpha R + \beta R^n$, leads to a boundary term which vanishes provided $K_{ij}$ - the velocity in disguise, is kept fixed at the boundary. Mannheim \cite{M} and Mannheim-Davidson \cite{MD} found no trouble regarding quantization holding position and velocity fixed at the end points, and apparently there is no problem as such. However, there are at least four reasons why, instead of $K_{ij}$ one should fix the scalar curvature $R$ at the boundary, and supplement the action with a boundary term, $\Sigma = 2\beta\int Kf'(R) \sqrt {h}d^3 x$, in addition to the GHY term. Firstly, the above boundary term reproduces the expected ADM energy upon passing to the Hamiltonian formalism \cite{DH}. Secondly, fixing $K_{ij}$ at the boundary, one looses the most cherished GHY term, which is identified with the entropy of a black hole. There is practically no physical interpretation available, why the concept of entropy of a black hole should get lost in strong gravity, and reappear when gravity is weak, particularly, when the correct expression of entropy of a Schwarzschild black hole has been found in the semiclassical limit \cite{DH}. Next, $f(R)$ gravity has been identified with scalar-tensor equivalent form in Jordan's frame of reference, under redefinition of $f'(R) = \Phi$ and $R = U_{,\Phi}$, where prime denotes derivative with respect to the Ricci scalar $R$, or in Einstein's frame of reference, under conformal transformation $\tilde{g}_{\mu\nu} = f'(R) g_{\mu\nu} = e^{2\omega} g_{\mu\nu}$, where the conformal factor $\omega$ is related to an effective scalar field $\tilde\phi$ by the relation $\omega = \sqrt{k\over 6}\tilde\phi$. Under variation, field equations are obtained keeping $\Phi$ fixed in Jordan frame and $\tilde\phi$ fixed in Einstein's frame, at the end points. This is equivalent to fix $R$ at the end points. This means scalar-tensor equivalence of higher order theory of gravity may be established, only under the condition, $\delta{R}|_{\partial\mathcal{V}} = 0$, which requires the supplementary boundary term, noted above. Finally, since $K_{ij}$ is the basic variable and also fixed at the boundary, it attains the same status as $g_{\mu\nu}$. So, there is no reason as to why the action should not be varied with respect to $K_{ij}$ too. Nevertheless, classical solution in the process, is restricted to just one possibility or even worse, to zero possibility \cite{Paddy}. Therefore, we suggest that instead of fixing $K_{ij}$, the Ricci scalar $R$ should be kept fixed at the boundary i.e. $\delta R|_{\partial\mathcal{V}}=0$, and the action should be supplemented by an additional boundary term in the form, $\Sigma = 2\beta\int Kf'(R) \sqrt {h}d^3 x$. Now, since $(h_{ij}, K_{ij})$ should be treated as basic variables for higher order theory of gravity, so in the Robertson-Walker metric (say for example)
\be\label{RW} ds^2 = -dt^2 + a^2(t)\Big[{dr^2\over 1-kr^2} + r^2(d\theta^2 + sin^2\theta d\phi^2)\Big]\ee
\bes h_{ij} = a^2 = z\; (\mathrm{say}),\;\;K_{ij} \propto 2 a\dot a = \dot z,\\& R = 6\left({\ddot a\over a} + {\dot a^2\over a^2} + {k\over a^2}\right) = 6\left({\ddot z\over 2z} + {k\over z}\right),\end{split}\ee
essentially the velocity ($\dot z$) is treated as an additional basic variable, while ($\delta R|_{\partial\mathcal{V}}=0$) implies acceleration ($\ddot z$) is kept fixed at the boundary in addition to $h_{ij} = z$. Therefore, we propose that in order to treat all higher order theories on the same footing, acceleration should be kept fixed at the boundary instead of the velocity. In section 2, we shall show in the context of Pais-Uhlenbeck action \cite{Pais}, how Ostrogradsi's \cite{Ostro} and Dirac's technique \cite{Dirac} fail to associate such proposition. Further, we show that under above proposition, Horowitz's technique \cite{Horo}, is not able to take into account the boundary terms in general. In this context, we have modified Horowitz's technique to construct a viable Hamiltonian formulation of a modified Pais-Uhlenbeck oscillator action and followed the procedure for higher order theory of gravity. \\

Canonical formulation of higher degree theory is yet another problem, which has not been resolved uniquely as yet. If the Lagrangian contains velocities beyond quadratic, momentum does not appear linearly in velocities, resulting in multivalued Hamiltonian with cusps, usually called the branched Hamiltonian. This makes classical theory unpredictable, since at any time one can jump from one branch of the Hamiltonian to the other. It also does not allow standard canonical formulation of the theory. Further, as energy has to be an observable, the multivalued Hamiltonian does not allow momentum to ensure a complete set of commuting observable. Finally, it also suffers from the disease of having non-unitary time evolution of the quantum state. Related issues arise in cosmological models in extensions of Einstein gravity involving topological invariants and in theories of higher-curvature gravity \cite{Topo}. For example, higher order theories of gravity are often plagued with non-unitary time evolution in the weak energy limit when expanded perturbatively about the flat Minkowski background, although the general theory might be free from such disease. To get rid of this unpleasant situation, even if one constructs a theory with a particular combination (Gauss-Bonnet) of the curvature invariant terms, the resulting Lanczos-Lovelock theory of gravity \cite{LL} is plagued with the problem of branched Hamiltonian. Such unpleasant issue arising out of branched Hamiltonian was addressed long ago \cite{T}. Starting from a toy model,

\be\label{A3} A_2 = \int \Big[{1\over 4}\dot q^4 - {1\over 2}\beta\dot q^2\Big]dt,\ee
Hennaux, Teitelboim and Zanelli \cite{T} had shown that in the path integral formalism one can associate a perfectly smooth quantum theory which possesses a clear operator interpretation and a smooth, deterministic, classical limit. Nevertheless, it puts up question regarding the canonical quantization scheme. Further, such formalism can't be extended in a more complicated theory beyond the toy model (\ref{A3}). The same issue has also been addressed by several other authors in the recent years \cite{S, Zhao, CH, Ebra}. However, in order to solve the problem, they had to tinker with some fundamental aspect viz., loosing the Heaviside function to obtain manifestly hermitian convolution together with the usual Heisenberg commutation relations \cite{S}, sacrificing the Darboux coordinate to parametrize the phase space \cite{Zhao},  and the standard Legendre transformation \cite{CH}. Additionally, the Hamiltonians obtained for the same above toy model following the above two prescriptions \cite{CH} and \cite{
Ebra} differ considerably, and not related through canonical transformation \cite{SRSA}. Therefore none of these techniques are fully developed or rigorous. However, one can bypass the problem by incorporating higher order term. For example, if action (\ref{A3}) is associated with a higher order term in the form

\be \label{A4} A_3 = \int \Big[\alpha\ddot q^2 + {1\over 4}\dot q^4 - {1\over 2}\beta\dot q^2\Big]dt,\ee
the problem associated with branching is resolved uniquely, in the process of canonical formulation of higher order theory. This has been demonstrated recently in the context of Lanczos-Lovelock gravity \cite{SRSA}. \\

In a nutshell, in the present manuscript we have first placed a proposition that in order to treat all higher order theories (including gravitation) on the same footing, one should fix acceleration instead of velocity at the end points. Once such proposition appears tenable, then neither Ostrogradski's, nor Dirac's technique can successfully handle the situation towards canonical formulation of Pais-Uhlenbeck fourth order oscillator action \cite{Pais}, which was presented by Mannheim and Davidson \cite{new}, along with other higher order theories. This we demonstrate in section (2.1). On the contrary, incorporating such proposition in Horowitz's technique it is found that the technique is unable to handle the counter terms in general. This has been demonstrated in section (2.2). In this context, we present a rigorous scheme for canonical formulation, by modifying Horowitz's method in section (2.3). This technique treats gravitational and non-gravitational higher order theories on the same footing, removing the pathologies discussed. Although, the Hamiltonian found in the process is canonically related to the ones found using other techniques for the actions under consideration, in the appendix we have demonstrated that for more general action, the Hamiltonian may be quite different and is not related to the others under canonical transformation. Second, we demonstrate in section (3) that it is possible to regulate the pathology of branching, by associating higher order term in higher degree theory.\\

\section{Fourth order oscillator}

Since our aim is to explore the problem associated with higher order theory, meaning that the Lagrangian is a function of at least second derivative of a field variable in the form ${\mathcal L = \mathcal L}(q_i, \dot q_i, \ddot q_i, \ddot q_i^2, ...)$ and the resulting Euler-Lagrange equation contains fourth or higher order terms, let us take up fourth order oscillator equation of motion, viz.

\be\label{p1} \stackrel{....}q+(w_1^2+w_2^2)\ddot q + w_1 w_2 q =0,\ee
which is supposed to be found from the Pais-Uhlenbeck action \cite{Pais}

\be \label{p2} S_1 = {\gamma\over 2}\int[\ddot q^2 - (w_1^2 + w_2^2)\dot q^2 + w_1w_2 q^2]dt.\ee
Canonical formulation of the above action following Dirac's constraint analysis \cite{Dirac} was found by
Mannheim and Davidson \cite{new} as

\be\label{H1} H_1 = {p_y^2\over 2\gamma} + y p_q + {\gamma\over 2}(w_1^2 + w_2^2) y^2 - {\gamma\over 2}w_1w_2 q^2,\ee
where, $y = \dot q$ and $p_y$ is the canonically conjugate momentum. In fact, it is trivial to show that the same could have been obtained following Ostrogradski's technique \cite{Ostro} also. However, under variation of the above action (\ref{p2}) one obtains

\be\label{p3} \delta S_1 = \gamma\int[\stackrel{....}q+(w_1^2+w_2^2)\ddot q + w_1w_2 q]\delta q dt +\gamma\ddot q\delta\dot q ,\ee
using the condition $\delta q|_ {\mathrm{end points}} = 0$. Now, to perform canonical analysis, an additional degree of freedom is required, which is $y = \dot q$. Therefore, to match boundary data with Cauchy initial data, yet another quantity is required to be fixed at the end points. Two options are now open. One is to set $\delta\dot q|_{endpoint} = 0$ and the other is to set $\delta\ddot q|_{endpoint} = 0$. For the second choice, the action (\ref{p2}) should be  supplemented with an additional term, so that it reads

\bes \label{p4} S_1 = {\gamma\over 2}\int[\ddot q^2 - (w_1^2 + w_2^2)\dot q^2 + w_1w_2 q^2]dt - \sigma_1\\&
= \int L_1 dt - \sigma_1\end{split}\ee
where, $\sigma_1 = \gamma\dot q\ddot q$. In Ostrogradski's \cite{Ostro} (together with its generalization developed for non-singular gravitational Lagrangian by Buchbinder-Lyakovich and Karataeva \cite{BL} following Dirac's constrained analysis) and of course in Dirac's \cite{Dirac} methods, first option is considered which apparently resolves the issue. But then, some problems appear. To better understand, let us consider the following oscillator action

\be\label{1}
S_2=\int \left[{k\over 2}\dot q^2\ddot q - {A\over 2}\dot q^2 + C U(q)\right]dt,\ee
which, despite the presence of $\ddot q$ term in the action, is not a higher order theory, since it does not produce fourth or higher order equation of motion. Clearly the first term is a total derivative term, which has been kept deliberately to administer the issue under discussion. Under variation (setting $\delta q = 0$ at end points) one obtains

\be\label{2}
\delta S_2 = \int (A\ddot q + C U')\delta q dt + {k\over 2}\dot q^2 \delta \dot q|_\mathrm{endpoints}, \ee
which produces the equation of motion for ordinary harmonic oscillator (in the above, prime stands for derivative with respect to $q$), if one can regulate the end point data. With $\delta \dot q = 0$ at the end points, the endpoint data exceeds Cauchy initial data. So, the only option is to find a counter term for action (\ref{1}), so that the boundary term gets cancelled. The complete action then reads

\be\label{3}
S_2=\int \left[{k\over 2}\dot q^2\ddot q - {A\over 2}\dot q^2 + C U(q)\right]dt - \sigma_2,\ee
where, $\sigma_2 = {k\over 6}\dot q^3$. Clearly, under integration by parts, $\ddot q$ term gets cancelled with the counter term, and the action takes the form of usual oscillator action, despite the presence of second derivative term in the action. Thus, the need for a boundary term is realized, through the simple example cited above. Let us now modify action (\ref{1}) by introducing $\ddot q^2$ term, so that it becomes the Pais-Uhlenbeck fourth order oscillator action with a total derivative term in the form

\be \label{4} S_3=\int \left[{B\over 2} \ddot q^2 + {k\over 2}\dot q^2\stackrel{..} q - {A\over 2}\dot q^2 + C U(q)\right]dt.\ee
Under variation one obtains

\bes\label{5}
\delta S_3 = \\&\int (B\stackrel{....}q + A\ddot q + C U')\delta q dt+ \Big({k\over 2}\dot q^2
+ B\ddot q\Big)\delta\dot q|_\mathrm{endpoints}\\&- (B\stackrel{...}q + A \dot q)\delta q|_\mathrm{endpoints}.\end{split}\ee
Now if one fixes both $q$ and $\dot q$ at the endpoints, boundary terms vanishes and there is no need for a counter term $\sigma_2$. Appropriate field equations are retrieved and so apparently there is no problem. However, there is no reason as to why the counter term $\sigma_2$ disappears, when higher order theory is considered. Further, for $B = 0$, the counter term $\sigma_2$ being absent, field equations are found only under the condition $q$ and $\dot q$ fixed at the end points, and in the process, boundary data exceeds Cauchy data.  \footnote{One would then resolve the issue of counter term appearing in lower order theory by adding a higher order term and then setting the higher order term to zero. In this way one can get rid of GHY term too. Clearly this is illegal.}. This is the same situation that one encounters in higher order theory of gravity, as mentioned in the introduction. Therefore, fixing both $q$ and $\dot q$ at endpoints lacks mathematical rigor.\\

On the contrary, if one chooses a counter term in the form $\sigma_3 = B\dot q\ddot q + {k\over 6}\dot q^3$ and set $\delta q = 0 =\delta \ddot q$ at the end points, then the complete action reads

\be \label{6} S_3=\int \left[{B\over 2} \ddot q^2 + {k\over 2}\dot q^2\stackrel{..} q - {A\over 2}\dot q^2 + C U(q)\right]dt - \sigma_3.\ee
Cauchy initial data now matches the boundary data and also, in the absence of higher order term ($B = 0$), the counter term for the harmonic oscillator action under present consideration (\ref{1}), is retrieved. Hence we propose that one should fix acceleration and not the velocity at the end points for fourth order theory and in the process, gravity is treated on the same footing as other higher order theories.

\subsection{Canonical formulation of Pais-Uhlenbeck action}

In view of preceding discussions, if the proposition of fixing acceleration at the endpoints instead of the velocity appears tenable, then obviously, Pais-Uhlenbeck action (\ref{p2}) should be replaced by (\ref{p4}), which contains a counter term. The question is how shall we proceed for canonical formulation of the said action, handling the counter term appropriately? One option is to take derivative of $\sigma_1$ and to enter it in the Lagrangian. In the process the action reads,

\be \label{p5}S_1 = {\gamma\over 2}\int[-\ddot q^2 -2 \dot q\stackrel{...}q - (w_1^2 + w_2^2)\dot q^2 + w_1w_2 q^2]dt.\ee
Under variation, action (\ref{p5}) produces appropriate equation of motion (\ref{p1}), setting $\delta q = 0 = \delta\ddot q$, at the end points. But, despite the fact that the above action produces fourth order Euler-Lagrange equation of motion, the appearance of third derivative term in the action functional requires yet another additional degree of freedom $x = \ddot q$, in order to follow Ostrogradski's technique \cite{Ostro} or to make Dirac constrained analysis \cite{Dirac}. In the process, Cauchy data exceeds boundary data. So both the techniques fail to handle fourth order oscillator action, under the current proposition. The other option is to take the action (\ref{p4}) and to make use of an auxiliary variable $Q$, following Horowitz \cite{Horo}. Although, the prescription was initially proposed for gravitational action, but since it is applied in minisuperspace model only, so it works for other higher order theories as well. Although, the prescription was initially proposed for gravitational action, but since it is applied in minisuperspace model only, so it works for other higher order theories as well. Although, velocity is kept fixed at the endpoints in this formalism too, it is indeed possible modify the technique by fixing acceleration instead, at the end points in Horowitz's formalism. The prime issue of this scheme then reads: ``first, one should choose an auxiliary variable varying the action with respect to the highest derivative of the (field) variables appearing in the action. Second, cast the action in canonical form, so that in the process of integration by parts, counter term gets cancelled. One can now vary the action both with respect to the fundamental variable $q$ and an auxiliary variable $Q$. While variation of $q$ yields appropriate Euler-Lagrange equation, variation with respect to $Q$ returns its definition only. Finally, as the phase space structure of the Hamiltonian is found, the auxiliary variable $Q$ should be replaced by the basic variable $\dot q$, following appropriate canonical transformation". The important point to note is that, $\dot q$ is treated as basic variable, but it doesn't acquire the same status as $q$, since, neither it is required to vary the action with respect to $q$, nor it is fixed at the end points. With the above prescription, the auxiliary variable for action (\ref{p4}) therefore is,

\be \label{p6} Q = {\partial L_1\over \partial \ddot q} = \gamma \ddot q,\ee
in view of which the action (\ref{p4}) may then be judiciously expressed as
\be S_1 = \int\Big[Q\ddot q  - {Q^2\over 2\gamma} - {\gamma\over 2}(w_1^2 + w_2^2)\dot q^2 + {\gamma\over 2}w_1w_2 q^2\Big] dt  - \sigma_1\ee
so that, under integration by parts, the boundary term $\sigma_1$ gets cancelled and one is left with the following canonical action
\be S_1 = \int\left[-{\dot Q\dot q} - {Q^2\over 2\gamma} - {\gamma\over 2}(w_1^2 + w_2^2)\dot q^2 + {\gamma\over 2}w_1w_2 q^2\right] dt.\ee
$Q$ variation equation returns the definition of $Q$ given in (\ref{p6}) and $q$ variation equation returns the fourth order oscillator equation of motion (\ref{p1}). The Hamiltonian reads
\be H_1 = -p_q p_Q + {\gamma\over 2}(w_1^2 + w_2^2) p_Q^2 + {Q^2\over 2\gamma} - {\gamma\over 2}w_1w_2 q^2.\ee
Now, phase-space structure of the Hamiltonian should be expressed in terms of the basic variables $q$ and $y = \dot q$ instead of $q~\mathrm{and}~ Q$, as mentioned. This may be achieved by the following canonical transformation
\be Q = {\partial L_1\over \partial \ddot q} = {\partial L_1\over \partial \dot y} = p_y\;\;\mathrm{and}\;\;p_Q = -y.\ee
Hence the final form of the Hamiltonian is
\be\label{H1} H_1 = {p_y^2\over 2\gamma} + y p_q + {\gamma\over 2}(w_1^2 + w_2^2) y^2 - {\gamma\over 2}w_1w_2 q^2.\ee
Note that this is the same Hamiltonian (\ref{H1}) obtained by Mannheim and Davidson \cite{new} following Dirac's constrained analysis. In fact, all the three formalisms discussed here, viz. Ostrogradski's \cite{Ostro}, Dirac's \cite{Dirac} and Horowitz's \cite{Horo}, produce the same and correct phase space Hamiltonian for the Pais-Uhlenbeck oscillator action. However, while Ostrogradski's and Dirac's formalism keep velocity at the endpoints fixed, it is indeed possible to fix acceleration at the end points in Horowitz's formalism. However, such a modified version of Horowitz's formalism is found to be diseased from certain pathology, which we expatiate underneath.

\subsection{Pathology appearing in Horowitz's formalism}

Although, Horowitz's formalism produces correct phase-space Hamiltonian, keeping velocity fix at the end points, it is indeed possible to fix acceleration at the end points too. However, such modification often does not guarantee to account for the supplementary boundary terms correctly. Let us consider action (\ref{6}) to explore the situation. This action produces fourth order oscillator equation of motion provided, endpoint data are taken care of, appropriately. Plugging in the auxiliary variable following Horowitz's prescription \cite{Horo},
\be Q = B\ddot q + {k\over 2}\dot q^2,\ee
judiciously into the action (\ref{6}) as,
\bes S_3 = \\&\int \Big[Q\ddot q - {1\over 2B}\Big(Q - {k\over 2}\dot q^2\Big)^2 -{A\over 2}\dot q^2
+ CU(q) \Big]dt\\& - \sigma_3,\end{split}\ee
it may be cast in the following canonical form under integration by parts
\bes S_3 = \\&\int \Big[-\dot Q\dot q - {1\over 2B}\Big(Q - {k\over 2}\dot q^2\Big)^2 -{A\over 2}\dot q^2
+ CU(q)\Big]dt\\& + Q\dot q - \sigma_3.\end{split}\ee
One can now trivially check that, $Q\dot q = B\dot q\ddot q + {k\over 2}\dot q^2 \ddot q$ does not cancel the counter term $\sigma_3$. The situation worsen for a coupled action in the form

\bes S = \gamma\int\Big[{1\over 2}\ddot q_1^2 - {1\over 2}A\dot q_1^2 + C U(q_1) + \lambda(q_2)\dot q_1^n\ddot q_1 \\&+ {1\over 2}\dot q_2^2
- V(q_2) \Big]dt,\end{split}\ee
which is encountered in dilatonically coupled Gauss-Bonnet action associated with higher order term (say, $R^2$) in $4$-dimensional Robertson-Walker minisuperspace model.\\

Finally, to apprehend the whole situation, let us cite a more concrete example in view of the following action
\be\label{7}
S_4 = {\gamma}\int \left[{1\over 2}\ddot q^2 + \ddot q^n\dddot q - {1\over 2}A\dot q^2 + C U(q)\right]dt.\ee
Under variation, one ends up with
\bes\delta S_4 = \gamma\int[\stackrel{....}q + A \ddot q + C U']\delta q dt \\&+ \gamma[\ddot q^n \delta \ddot q + \ddot q\delta \dot q
- \dddot q \delta q - A \dot q \delta q]\Big|_{\mathrm{endpoints}}.\end{split}\ee
Thus, the same fourth order oscillator equation of motion (\ref{p1}) is reproduced, under the choice $A = w_1^2 + w_2^2,~C = w_1w_2~\mathrm{and}~U(q) = {1\over 2} q^2$, with appropriate end point data. However, following Ostrogradski's or Dirac's prescription (setting $\delta q = 0 = \delta\dot q$ at endpoints), one can't retrieve the equation of motion, unless $\delta\ddot q|_{\mathrm{endpoints}} = 0$. But then, endpoint data exceeds Cauchy data. Otherwise, one would require an additional degree of freedom to handle the situation. Nevertheless, in that case, Cauchy data exceeds endpoint data \footnote{One might suggest that, set $q$ and $\dot q$ at end points and supplement the action with a counter term $\gamma {\ddot q^n\over n+1}$, and thus the problem is resolved. But then the question arises, ``at any stage if we require a counter term, then why shall we sacrifice the GBY term?"}. On the contrary, the action may be supplemented by,
\be\label{sig5}\sigma_4 = \gamma\left(\dot q\ddot q + {\ddot q^{n+1}\over n+1}\right),\ee
so that setting $\delta q = 0 = \delta\ddot q$ at the end points, it reads,
\bes \label{8}
S_4 = {\gamma}\int \left[{1\over 2}\ddot q^2 + \ddot q^n\dddot q - {1\over 2}A\dot q^2 + C U(q)\right]dt - \sigma_4\\&
= \int L_4 dt - \sigma_4.\end{split}\ee
Now, following Horowitz's prescription, the auxiliary variable $Q$ is defined as
\be Q = {\partial L_4\over \partial\dddot q} = \gamma \ddot q^n.\ee
It now turns extremely difficult, if not impossible to cast the action (\ref{8}) in canonical form. As a result it is not known if the total derivative term cancels the supplementary boundary terms.

\subsection{A legitimate Canonical formulation}

At this end, we understand that, for higher order theory under consideration, if our proposition that $\delta q = 0 = \delta\ddot q$ at the endpoints, appears tenable, then neither Ostrogradski's nor Dirac's canonical scheme works. On the contrary, although Horowitz's formalism appears mathematically rigorous, it can't handle the supplementary counter terms, in general. However, the pathology appearing in Horowitz's formalism disappears if the action is integrated by parts, prior to the introduction of auxiliary variable. For example, the last action (\ref{8}) under integration by parts yields

\be\label{9}S_4 = {\gamma}\int \left[{1\over 2}\ddot q^2 - {1\over 2}A\dot q^2 + C U(q)\right]dt-\gamma\dot q\ddot q.\ee
The above action (\ref{9}) takes the form of Pais-Ulhenbeck action (\ref{p4}). Canonical formulation then follows as depicted in section (2.1). The same technique resolves the problem of all higher order actions as well.\\

In the case of gravity however, choice of appropriate variable is a precursor to handle the associated boundary terms \cite{Sanyal, Sanyal1, Sanyal2, Sanyal3}. As an example let us consider the following action

\bes\label{grav} A_4 = \int\left[{R\over 16\pi G} + \beta R^2\right]\sqrt{-g} d^4 x\\&
+ {1\over 8\pi G}\int K\sqrt{h} d^3 x + 4\beta\int {^4R}K\sqrt{h} d^3x,\end{split}\ee
which, in the Robertson-Walker minisuperspace (\ref{RW}) reads

\bes A_4 = \int\left[{3\over 8\pi G}(a^2 \ddot a+a\dot a^2 + k a) + 36\beta\bigg(a\ddot a^2 \right.\\&
\left.+ 2 \dot a^2 \ddot a+2k\ddot a+{(\dot a^2 + k)^2\over a}\bigg)\right]d^3x dt\\&
-{3\over 8\pi G}\int a^2 \dot a d^3x - 72\beta\int\left(a\dot a\ddot a+\dot a^3+k\dot a\right) d^3x.\end{split} \ee
Under integration by parts one ends up with
\bes A_4 = \int\left[{3\over 8\pi G}(-a\dot a^2 + k a) + 36\beta\bigg(a\ddot a^2\right.\\&
\left.+{(\dot a^2 + k)^2\over a}\bigg)\right]d^3 x dt - 72\beta\int\left(a\dot a\ddot a - {2\over 3}\dot a^3\right) d^3x\\&
=\int\mathcal{L} dt- 72\beta\int\left(a\dot a\ddot a - {2\over 3}\dot a^3\right) d^3x. \end{split}\ee
Now introducing the auxiliary variable
\be Q = {\partial \mathcal{L}\over \partial\ddot a}=72\beta a\ddot a, \ee
judiciously in the above action

\bes A_4 = \int\left[{3\over 8\pi G}(-a\dot a^2 + k a) + Q\ddot a - {Q^2\over 144\beta a}\right.\\&
\left.+36\beta{(\dot a^2 + k)^2\over a}\right]d^3 x dt - 72\beta\int\left(a\dot a\ddot a - {2\over 3}\dot a^3\right) d^3x, \end{split}\ee
it takes the following canonical form, under integration by parts
\bes A_4 = \int\left[{3\over 8\pi G}(-a\dot a^2 + k a) - \dot Q\dot a - {Q^2\over 144\beta a}\right.\\&
\left.+36\beta{(\dot a^2 + k)^2\over a}\right]d^3 x dt +48\dot a^3 d^3x. \end{split}\ee
Nevertheless, one retains an additional boundary term. This is due to bad choice of co-ordinate, as we know, the basic variable is, $h_{ij} = a^2$ and not `$a$'. One can observe that under the choice $h_{ij} = a^2 = z$, the action (\ref{grav}) may be cast as

\bes A_4 = \int\left[{3\over 8\pi G}\left({\sqrt z\ddot z\over 2}+ k\sqrt z\right)+36\beta\left({\ddot z^2\over 4\sqrt z}+{k\ddot z\over \sqrt z}\right.\right.\\&
\left.\left.+{k^2\sqrt z}\right)\right]d^3x dt-{3\over 16\pi G}\sqrt z\dot z- 36 \beta\left({\dot z\ddot z\over 2\sqrt z}+{k\dot z\over\sqrt z}\right),\end{split}\ee
which under integration by parts, becomes

\bes A_4 = \int\left[{3\over 8\pi G}\left(-{\dot z^2\over 4\sqrt z}+ k\sqrt z\right)+36\beta\left({\ddot z^2\over 4\sqrt z}\right.\right.\\&
\left.\left.+{k\dot z^2\over z^{3\over 2}}+{k^2\sqrt z}\right)\right]dt- 36 \beta\left({\dot z\ddot z\over 2\sqrt z}\right).\end{split}\ee
Now, plugging in the auxiliary variable
\be Q = {\partial \mathcal{L}\over \partial \ddot z} = 18{\ddot z\sqrt z},\ee
the action reads
\bes A_4 = \int\left[{3\over 8\pi G}\left(-{\dot z^2\over 4\sqrt z}+ k\sqrt z\right)+Q\ddot z - {Q^2\sqrt z\over 36}\right.\\&
\left.+36\beta\left({k\dot z^2\over z^{3\over 2}}+{k^2\sqrt z}\right)\right]dt- 36 \beta\left({\dot z\ddot z\over 2\sqrt z}\right),\end{split}\ee
which under integration by parts yields the following canonical form
\bes A_4 = \int\left[{3\over 8\pi G}\left(-{\dot z^2\over 4\sqrt z}+ k\sqrt z\right)-\dot Q\dot z - {Q^2\sqrt z\over 36}\right.\\&
\left.+36\beta\left({k\dot z^2\over z^{3\over 2}}+{k^2\sqrt z}\right)\right]dt.\end{split}\ee
In the process, the problem with the boundary term disappears and one can now proceed to cast, what we call - the legitimate canonical formulation of higher-order theory, in which all the higher order theories (including gravitation) may be treated on the same footing.\\

\section{Resolving the issue of branched Hamiltonian}

Since, we have been able to construct a viable canonical formulation, that fits both the modified fourth order oscillator action and gravity as well, let us turn our attention to resolve the issue of branched Hamiltonian. We have already mentioned that if a Lagrangian contains velocities with degree greater than two, the resulting Hamiltonian is multivalued, and the standard Legendre transformation to cast the action in canonical does not apply in general. Despite attempts, the problem associated with branched Hamiltonian has not been resolved uniquely. However, as already mentioned, such situation often appears with higher order theory. For example, in the context of gravity, higher order curvature invariants ($R^2, R_{\mu\nu}R^{\mu\nu}$ etc.) contain terms with higher degree in the field variables. We shall show that pathology of branching may be removed, if the action accompanies higher order term in addition, as presented in action (\ref{A4}).

\subsection{Generalized Pais-Uhlenbeck action}

Note that in the presence of an appropriate potential term, action (\ref{A4}) leads to Pais-Uhlenbeck oscillator action (\ref{p2}) accompanying a higher degree ($\dot q^4$) term. Therefore, for the sake of demonstration, let us take the modified version of the forth order oscillator action (\ref{8}), by associating with it a higher degree term, so that the Lagrangian ${\mathcal L = \mathcal L}(q_i, \dot q_i, \ddot q_i, \ddot q_i^2, ..., \dot q_i^4)$,  and dub it as the generalized fourth order oscillator action. Such an action reads
\be S_5 = \gamma\int\left[\frac{1}{2}\ddot q^2 + \dot q^n \ddot q - {A\over2}\dot q^2 + C U(q) + {D\over 4}\dot q^4\right]dt -\sigma_4.\ee
In the absence of higher order ($\ddot q^2$) term, the above action suffers from the pathology of branching. However, in the presence of higher order term, integrating by parts yields
\bes S_5 = {\gamma}\int \left[{1\over 2}\ddot q^2 - {1\over 2}A\dot q^2 + C U(q) + {D\over 4}\dot q^4\right]dt-\gamma\dot q\ddot q\\&
= \int\mathcal{L}_5dt -\gamma\dot q\ddot q.\end{split}\ee
Introducing the auxiliary variable
\be Q = {\partial \mathcal{L}_5\over\partial\ddot q} = \gamma\ddot q\ee
into the action as before and under integration by parts, one obtains

\be S_5 = \int \left[-\dot Q\dot q - {Q^2\over 2\gamma} - {\gamma\over 2}A\dot q^2 + \gamma C U(q) + {\gamma D\over 4}\dot q^4\right]dt.\ee
The phase space structure of the Hamiltonian may then be obtained at-ease as,
\be H = -p_Q p_q + {Q^2\over 2\gamma} + {\gamma A\over 2}p_Q^2  - \gamma C U(q) - {\gamma D\over 4}p_Q^4.\ee
Now replacing $Q$ and $p_Q$ by basic variables under canonical transformations, $Q = p_y$ and $p_Q = -y$, where, $y = \dot q$, as before, the Hamiltonian may be expressed in its final form as
\be H = {p_y^2\over 2\gamma} + y p_q +\gamma \Big(\frac{A}{2} y^2  - C U(q) - {D\over 4} y^4\Big).\ee
Clearly, the velocities appear in the form of potential and the pathology appearing from branching is resolved.

\subsection{Higher order theory of gravity}

For the sake of completeness, let us turn our attention to higher order theory of gravity, yet again. Canonical formulation of gravity requires a $3+1$ decomposition in $4$-dimension, under which $h_{ij}$ is treated as basic variable. In higher order theory of gravity, the extrinsic curvature tensor $K_{ij}$ is treated as an additional variable, which essentially is the first time derivative of $h_{ij}$. Further, higher order theory of gravity requires to fix $R$ at the boundary in addition, which in principle may be expressed in terms of $h_{ij}$ and its second time derivative, viz. $\ddot h_{ij}$. To better understand, let us take Robertson-Walker minisuperspace model (\ref{RW}), as an example. Apart from the $3$-volume factor, since $h_{ij} = a^2$, so instead of the scale factor $a$, one should treat $z = a^2$ as basic variable. Automatically, the additional variable is $\dot z = 2 a\dot a \propto K_{ij}$. Therefore to reconcile boundary conditions, in addition to $z$, $\ddot z$ should be kept fixed at the boundary. Now,
\be R = 6\left({\ddot a \over a} + {\dot a^2\over a^2} + {k\over a^2}\right) = 3\left({\ddot z\over z} + 2{k\over z}\right).\ee
Therefore, keeping $\ddot z$ fixed is analogous to keep $R$ fixed at the boundary, as already mentioned in the introduction \footnote{In $D$ dimension, $R = (D-1)\left[2{\ddot a\over a} + (D-2){\dot a^2\over a^2}\right] + 6{k\over a^2} = 4{D-1\over D}{\ddot z\over z} + 6{k\over z^{4\over D}}$, under the choice, $z = a^{D\over 2}$. Particularly, in a synchronous frame $ds^2 = -dt^2 + h_{\alpha\beta}d x^{\alpha}d x^{\beta}$, it is always possible to express $R$ in terms of $h_{ij}~\mathrm{and}~\ddot h_{ij}$.}. Hence there is indeed no contradiction with higher order theory treated so far, since the choice of basic variables and boundary conditions reconcile.\\

Next, it is required to see how multi-valued Hamiltonian appears in the theory of gravity, and may be handled taking into account higher order curvature invariant terms. It is well known that gauge invariant divergences make general theory of relativity non-renormalizable and a renormalizable theory of gravity in $4$-dimension requires to modify Einstein-Hilbert action by incorporating curvature squared terms in the form \cite{stelle}
\be A_5 =\int\sqrt{-g}d^4x\left(\frac{R - 2\Lambda}{2\kappa} + \beta R^2 + \gamma R_{\mu\nu}R^{\mu\nu} \right),\ee
However, analysis of linearized radiation reveals the presence of five massive spin-2 particles whose excitations are negative definite and therefore are ghosts. These ghosts destroy the unitarity and so no sensible physical interpretation of such a theory exists. In the absence of $R_{\mu\nu}R^{\mu\nu}$ term ghosts disappear, but the theory becomes non-renormalizable. Although scalar curvature square term $R^2$ is not responsible for the appearance of ghosts, yet it is a general practice to cast the gravitational action with a particular combination of higher order curvature invariant terms so that fourth derivative terms disappear from the field equations. Gauss-Bonnet term ($R^2 - 4R_{\mu\nu}R^{\mu\nu} + R_{\alpha\beta\mu\nu}R^{\alpha\beta\mu\nu}$) was recalled for the purpose. As it is topological invariant in $4$-dimensions, Lanczos-Lovelock gravity \cite{LL} was therefore constructed, which is realizable in dimensions higher than four. Nevertheless, it was found that the action contains higher degree term, leading to branched Hamiltonian and so the theory again lacks unitary time evolution of quantum states. This issue was resolved by modifying Lanczos-Lovelock gravity under the addition of a scalar curvature square term ($R^2$) \cite{SRSA}. In the following, we therefore take up yet another example in $4$-dimension. \\

The combination of higher order terms in the form $\mathcal{R}_2=R-\sqrt{3(4R_{\mu\nu} R^{\mu\nu}-R^2)}$ also produces second order field equations, although in Robertson-Walker mini-superspace (\ref{RW}) only. Hence, it might appear that the following action
\bes\label{bh.1}
A_6=\int\sqrt{-g}d^4x\left({R\over 2\kappa}  + B_2 \mathcal{R}_2^2- \frac{1}{2}\phi_{,\mu}\phi^{,\mu} - V(\phi)\right)\\&
+ {1\over \kappa} \int d^3 x {\sqrt h} K,
\end{split}\ee
should be free from all pathologies at least in Robertson-Walker minisuperspace metric (\ref{RW}) and one does not have to bother about the boundary term $d^3x {\sqrt h} K f'(R) $ along with additional boundary condition $(\delta R|_{\partial {\mathcal V}} =0)$ appearing from $f(R)$ theory of gravity, other than Gibbons-Hawking-York term \cite{Boun}. Further, note that since in isotropic space-time, $(R_{\mu\nu} R^{\mu\nu} - {R^2\over 3})$ is a total derivative term, so $R_{\mu\nu}R^{\mu\nu}$ is usually replaced safely by $R^2$ term. However, as mentioned, the theory is not renormalizable in the absence of $R_{\mu\nu}^2$ term. But here, the presence of such a combination within the square root does not allow such replacement and hence both the terms are retained as such. Now, in terms of the basic variable $(h_{ij} = a^ 2 = z)$, the Ricci scalar and the higher order curvature invariant term read $R = 6({\ddot z\over 2 z} + {k\over z})$  and $\mathcal{R}_2=12\left(\frac{\dot z^2}{4z^2}+\frac{k}{z}\right)$
respectively. After getting rid of the GHY boundary term under integration by parts, the action (\ref{bh.1}) can therefore be expressed as
\bes\label{bh.2}
A_6=\int\left[{3\over \kappa}\left(-{\dot z^2\over 4\sqrt z} + {k\sqrt z}\right)\right.+\\& 144B_2\left(\frac{\dot z^2}{4z^\frac{5}{4}}+\frac{k}{z^{1\over 4}}\right)^2\left.+\left(\frac{1}{2}\dot\phi^2 - V(\phi)\right)z^{\frac{3}{2}}\right]dt.
\end{split}\ee
The presence of only first time derivative of the field variables clearly indicates that despite the presence of higher order curvature invariant terms, the field equations are only second order. However, the canonical momentum
\be p_z = -{3\over \kappa} {\dot z\over 2\sqrt z} + 144 B_2 \left({\dot z^3 \over 4 z^{5\over 2}} + {k\dot z\over  z^{3\over 2} }\right),\ee
involves $\dot z^3$ term, leading to the issue of branched Hamiltonian. Therefore, while perturbatively higher order terms  destroy unitarity, nonperturbatively, higher degree terms destroy the same. However, such a worse situation may be resolved by adding yet another curvature invariant term $\mathcal{R}_1^2$ where, $\mathcal{R}_1=R+\sqrt{3(4R_{\mu\nu} R^{\mu\nu}-R^2)}=12\left({\ddot z\over 2z}-{\dot z^2\over 4z^2}\right)$ to the action \eqref{bh.1}, so that the modified action reads
\bes\label{bh2.1}
A_7=\\&
\int\sqrt{-g}d^4x\left({R\over 2\kappa}  + B_1 \mathcal{R}_1^2+B_2\mathcal{R}_2^2- \frac{1}{2}\phi_{,\mu}\phi^{,\mu}\right.\\&
- V(\phi)\bigg)+\sigma+\Sigma,
\end{split}\ee
where, $\sigma$ is the GHY boundary term \cite{Boun} and $\Sigma = 4B_1\int {^4R}\sqrt h K d^3x$, appears under variation, with an additional condition that $\delta R = 0$ at the boundary. In fact, for $B_2 = B_1$, the above combination gives scalar curvature squared term $R^2 = \mathcal{R}_1^2+\mathcal{R}_2^2$ and the action simply reads
\bes\label{5.1} A_7 = \\&
\int \Big[{1\over 2\kappa} R + B_1 R^2 - \frac{1}{2}\phi_{,\mu}\phi^{,\mu} - V(\phi)\Big]\sqrt{-g}d^4x \\&
+ \sigma+\Sigma.\end{split}\ee
A legitimate canonical formulation of the above action in the absence of the scalar field exists in the literature taking lapse function into account \cite{Sanyal2, Sanyal3}. Here we briefly discuss the outcome, in the context of the issue of boundary term, canonical quantization and the additional boundary condition in connection with Noether symmetry.\\

\noindent
\textbf{The issue of boundary term:}\\

In Robertson-Walker space-time (\ref{RW}) the action (\ref{5.1}), under the choice ($B = 36 B_1$) reads
\bes\label{5.2}
 A_7 = \\&
 \int \Big[{3\over \kappa}\left({\sqrt z\ddot z\over 2} + {k\sqrt z}\right)+ B\left(\frac{\ddot z^2}{4\sqrt z} + \frac{k^2}{\sqrt z} + k\frac{\ddot z}{\sqrt z} \right)\\&
+\left(\frac{1}{2}\dot\phi^2 - V(\phi)\right)z^{\frac{3}{2}}\Big]dt + \sigma + \Sigma_1 + \Sigma_2.\end{split}\ee
In the above, the supplementary boundary term has been split as, $\Sigma = 4B_1\int({^4R})K\sqrt{h} d^{3} x= \Sigma_1 + \Sigma_2 =4B_1\int({^3R})K\sqrt{h} d^{3} x + 4B_1\int({^4R}-{^3R})K\sqrt{h} d^{3} x$. Under integration by parts, the surface terms $\sigma$ and $\Sigma_1$ get cancelled and the action takes the form,
\bes\label{5.4}A_6 = \int \left[{3\over \kappa}\left(-{\dot z^2\over 4\sqrt z} + {k\sqrt z}\right)+B\left(\frac{\ddot z^2}{4\sqrt z}+ \frac{k\dot z^2}{2z^{\frac{3}{2}}} \right.\right.\\&
\left.\left. +\frac{k^2}{\sqrt z}\right)+ \left(\frac{1}{2}\dot\phi^2 - V(\phi)\right)z^{\frac{3}{2}}\right]dt + \Sigma_2 \\&
= \int \mathcal{L}_7 dt + \Sigma_2,\end{split}\ee
which in view of the auxiliary variable
\be\label{5.5} Q = \frac{\partial \mathcal{L}_7}{\partial \ddot z} = {B\over 2}\frac{\ddot z}{\sqrt z},\ee
may be cast in the following canonical form
\bes\label{5.7}
A_7 = \int\Big[{3\over \kappa}\left(-{\dot z^2\over 4\sqrt z} + {k\sqrt z}\right)-\dot Q\dot z - \frac{Q^2\sqrt z}{B}\\&+ \frac{Bk\dot z^2}{2z^{\frac{3}{2}}}+\frac{Bk^2}{\sqrt z}+\Big(\frac{1}{2}\dot\phi^2 - V(\phi)\Big)z^{\frac{3}{2}}\Big]dt,\end{split}\ee
and in the process, the boundary term $\Sigma_2$ gets cancelled with the total derivative term. Therefore, boundary terms have been taken care of appropriately, in two stages. The canonical momenta are
\be\label{Mom} p_Q = -\dot z,\;p_z = -{3\over 2\kappa}{\dot z\over\sqrt z}-\dot Q +{Bk\dot z\over z^{3\over 2}},\;p_{\phi} = \dot \phi z^{3\over 2}.\ee

\noindent
\textbf{Canonical quantization:}\\

The phase-space structure of the Hamiltonian is,
\bes\label{5.20}
H={3\over 4\kappa}{p_Q^2\over \sqrt z}-{3\over \kappa} k\sqrt z-{B k p_Q^2\over2z^{3\over 2}}+{p_\phi^2\over 2z^{3\over 2}}-p_z p_Q\\&+{Q^2\sqrt z\over B}-{Bk^2\over \sqrt z}+V(\phi)z^{3\over 2}=0.\end{split}\ee
Now, in view of the definition of the auxiliary variable and its canonical conjugate momenta given in expressions (\ref{5.5}) and (\ref{Mom}) respectively, we are in a position to translate the Hamiltonian in terms of the basic variables $(h_{ij}, K_{ij})$. For the purpose, we replace $Q$ by $p_y$ and $p_Q$ by $-y$, where $y = \dot z$. The Hamiltonian therefore takes the form
\bes H = y p_z +{\sqrt z\over B}p_y^2 + {p_{\phi}^2\over 2z^{3\over 2}} + {3\over 4\kappa} {y^2 \over \sqrt z}-{{3\over \kappa} k\sqrt z}\\&+V z^{3\over 2}- {B k y^2\over 2 z^{3\over 2}} - {Bk^2\over\sqrt z},\end{split}\ee
which is constrained to vanish. Canonical quantization then leads to the following Schr\"odinger-like equation
\bes\label{5.21}
{i\hbar\over \sqrt z}{\partial\Psi\over\partial z}={i\hbar}{\partial\Psi\over\partial \eta}\\&
=-{\hbar^2 \over B }\left[{1\over y}{\partial^2\over\partial y^2}+ {n\over y^2}{\partial\over\partial y}\right]\Psi-{\hbar^2 \over 2\eta^{4\over 3}y}{\partial^2\Psi\over\partial \phi^2}\\&
+\left({3\over 4\kappa} {y \over  \eta^{2\over 3}}-{3\over \kappa}{k\over y}+{\eta^{2\over 3}\over y}V(\phi) - {B k y\over 2 \eta^{4\over 3}} - {Bk^2\over y \eta^{2\over 3}}\right)\Psi,\end{split}\ee
where, $\eta = z^{3\over 2} = a^3$ - the proper volume, acts as internal time parameter and $n$ is the operator ordering index. Clearly one ends up with a hermitian effective Hamiltonian operator with unitary time evolution of quantum states. In the process, the issue of branched Hamiltonian has also been resolved. It has been shown \cite{Sanyal2, Sanyal3} that the above quantum description leads to well-posed probabilistic interpretation, fixing the operator ordering index to $n = -1$. Further, the semiclassical wave-function has been found to be oscillatory, indicating that the region is classically allowed and is strongly peaked about a set of inflationary solutions admissible by the classical field equations.\\

\noindent
\textbf{Noether symmetry:}\\

First of all let us mention that in the process of canonical formulation of higher order theory following Ostrogradsi's and Dirac's techniques, it is not possible to explore Noether symmetry at least in the Lagrangian formalism. However, this may be expatiated under the present technique of canonical formulation. Noether symmetry for finding the form of the scalar potential corresponding to higher order theory of gravity in addition to non-minimally coupled scalar-tensor action already exists in the literature \cite{4}. It was there observed that the classical solutions of the field equations are obtained directly from the Noether equations. For the sake of simplicity, here we ignore the Einstein-Hilbert part of the action (\ref{5.1}), which is of-course ignorable in the very early universe. In the Noether symmetry approach, the lift vector $X$ acts as an infinitesimal generator of Noether symmetry in the tangent space $(z,\dot z, Q, \dot Q, \phi,\dot \phi)$, and is introduced as follows,

\be\label{lv} X = \alpha \frac{\partial }{\partial z}+\beta\frac{\partial}{\partial Q} + \gamma\frac{\partial }{\partial \phi}
+\dot\alpha \frac{\partial }{\partial\dot z}+ +\dot\alpha \frac{\partial }{\partial\dot Q} +\dot\beta\frac{\partial }{\partial\dot \phi}. \ee
The existence condition for symmetry, $\pounds_X L =X L = 0 $, then leads to the following master Noether equation, in view of the canonical action (\ref{5.7}),
\bes \label{5.8}\alpha\left[-\frac{Q^2}{2B\sqrt z}-\frac{3Bk\dot z^2}{4z^{\frac{5}{2}}}-\frac{Bk^2}{2z^{\frac{3}{2}}}+\frac{3}{2}\Big(\frac{\dot\phi^2}{2} -V\Big)\sqrt z\right]\\&
-\frac{2\beta}{B}Q\sqrt z
-\gamma z^{\frac{3}{2}}V'+\big(\alpha_{,z}\dot z+\alpha_{,Q}\dot Q+\alpha'\dot\phi\big)\Big(\frac{Bk\dot z}{z^{\frac{3}{2}}}\\&
 -\dot Q\Big)-\big(\beta_{,z}\dot z+\beta_{,Q}\dot Q+\beta'\dot\phi\big)\dot z+\big(\gamma_{,z}\dot z+\gamma_{,Q}\dot Q\\&
 +\gamma'\dot\phi\big)\dot \phi z^{\frac{3}{2}}=0.\end{split}\ee
Equating coefficients ($\dot z^2,\dot Q^2,\dot\phi^2, \dot z\dot Q, \dot Q\dot \phi,\dot\phi\dot z$) and the terms independent of time derivative to zero, as usual, we obtain following set of partial differential equations, viz.,
\bes\label{5.9} B k\big(z\alpha_{,z} -\frac{3}{4}\alpha\big) -z^{\frac{5}{2}}\beta_{,z}=0=
\alpha_{,Q} = 3\alpha +4 z \gamma'\\&
\alpha_{,z} + \beta_{,Q} = 0=
z^{\frac{3}{2}}\gamma_{,Q} -\alpha' =
 B k \alpha' + z^{\frac{3}{2}}\beta' - z^3 \gamma_{,z}\\&
\alpha\left[\frac{Q^2}{2 B \sqrt z}+\frac{Bk^2}{2z^{\frac{3}{2}}}+\frac{3}{2}\sqrt{z} V\right]+\frac{2\beta}{B}Q\sqrt z+z^{\frac{3}{2}}\gamma V'=0.\end{split}\ee
These set of equations may be solved for $k = 0$ to yield,
\bes\alpha = -\frac{4 l}{3}z;\;\beta  = {4 l\over 3} Q ;\;\gamma = l\phi\\&
V(\phi)= V_0\phi^2 + \frac{m^2}{4} ;\; Q = \frac{m\sqrt B}{2}\sqrt z,\end{split}\ee
where, $l$ and $m$ are the separation constants. In view of the definition of the auxiliary variable (\ref{5.5}), $R (= {3m\over \sqrt B})$ turns out to be a constant, leading to de-Sitter or anti-de Sitter solution depending on the separation constant $m$.
The conserved current
\be\label{5.18} {\mathcal{I}} = \frac{4l}{3}\dot Q z - {4l\over 3} Q\dot z + l \phi\dot\phi z^{\frac{3}{2}},\ee
satisfies the field equations and everything is well behaved. Since, Noether symmetry demands $R$ to be constant, it is trivially constant at the boundary. Thus, in view of Noether symmetry, $R^2$ gravity admits de-Sitter or anti de-Sitter solutions and nothing else. In a recent article \cite{we}, it was shown that Noether symmetry of higher order theory of gravity administers the same condition ($R$ = constant) both in isotropic and anisotropic space-times. Thus, in view of Noether symmetry, we get an answer to the question ``why higher order theory of gravity demands $R$ should remain fixed at the boundary?".

\section{Concluding remarks}

As mentioned in the introduction, there are several reasons which favour to keep the Ricci scalar ($R$) fixed at the boundary, for higher order theory of gravity, in addition to the metric ($g_{\mu\nu}$). Of particular importance is, scalar-tensor equivalence of higher order theory of gravity requires to fix $R$ at the boundary. This means keeping the acceleration fixed at the end points, in disguise. On the contrary, Ostrogradski's or Dirac's techniques for canonical formulation of fourth order theory, tacitly assumes velocity should be fixed, instead. In the process, not only GHY term, which is related to the entropy of a black hole gets lost, but also there exists a clear contradiction in handling different higher order theories.\\

To remove such contradiction and to treat different higher order theories on the same footing, we propose that acceleration, in addition to the coordinate should remain fixed at the endpoints. Such proposition, as mentioned, however is not compatible with Ostrogradski's and Dirac's techniques. Nevertheless, Horowitz's technique of associating an auxiliary variable, although originally developed for canonical formulation of higher order theory of gravity, appears to work in general with the said proposition. Unfortunately, with such proposition, Horowitz's formalism also often fails to handle supplementary boundary terms. In this connection, we have modified the technique and in the process, a one-to-one correspondence between canonical formulation of higher order oscillator action and gravity has been established. Further, in the appendix we have shown that the phase-space Hamiltonian constructed following such technique is different and is not related to the others under canonical transformation. The modified version is unique in this sense.\\

The presence of higher degree term in velocity leads to branched Hamiltonian and tells upon unitary time evolution of quantum states. Despite attempts, there does not exist a unique technique to resolve the issue. Under such circumstances, we have also demonstrated the fact that, the pathology due to the appearance of such multivalued Hamiltonian may be regulated if it appears with higher order theory. The technique was earlier applied to alleviate the pathology of branched Hamiltonian appearing in Lanczos-Lovelock gravity, by adding a scalar curvature squared term \cite{SRSA}. Here, the technique has been applied in a theory of gravitation, which appears to be free from all pathologies in connection with renomalizability and the appearance of ghost degree of freedom, of-course in isotropic and homogeneous cosmology. However, the problem has been resolved by adding a higher order term. The technique is well posed, leading to Schr\"odinger like equation with straight forward probabilistic interpretation, which is lacking in Wheeler-DeWitt equation. Semiclassical approximation had been found to be strongly peaked about classical inflationary solution \cite{Sanyal2, Sanyal3}.\\

In the process of canonical formulation following Ostrogradski's or Dirac's technique, the Lagrangian is left alone, and indeed there is no way to explore Noether symmetry at least in the Lagrangian formulation. Use of auxiliary variable on the contrary, is well suited for the purpose. Now, one way to resolve the issue of dark energy is to modify the left hand side of Einstein's equation by adding higher order curvature invariant terms. Usually, one chooses $F(R)$ by hand or following the reconstruction scheme. Recently, in an attempt to find a form of $F(R)$ by invoking Noether symmetry, both in isotropic and anisotropic space-time, it was observed that Noether symmetry demands $R$ must remain constant, resulting in de-Sitter/anti de-Sitter solutions \cite{we}. This at least gives a signal that nothing other than $R$ could be fixed at the boundary. Here again we observe that Noether symmetry requires $R = R_0$ - a constant, and as such administers the additional boundary condition $\delta R|_{\partial\mathcal{V}} = 0$.\\

\noindent
\texttt{{Acknowledgement:}} Authors would like to thank IUCAA, India, for their hospitality. Kaushik Sarkar would like to thank RGNF, UGC (India) for providing a fellowship.

\appendix
\section{Different phase-space Hamiltonian}
In the appendix, we cite an example to demonstrate that not all phase-space Hamiltonian corresponding to higher order theory, constructed out of different canonical formalisms are related through canonical transformation. Let us take the following higher order action,

\be A = \int\left[f(\phi)(\ddot a+a)^2 + {1\over 2}\dot a^2 + {1\over 2}\dot\phi^2 - V(\phi)\right]dt.\ee
Now, choosing $\dot a = x$, the above Lagrangian

\be L = (\dot x + a)^2 f(\phi)+ {x^2\over 2} + {\dot\phi^2\over 2} - V(\phi).\ee
becomes singular and so instead of Ostdtrogradski's technique, one should follow Dirac's constrained analysis, fixing $a$ and $\dot a = x$ at the boundary.
Introducing Lagrange multiplier $\lambda$, one then writes

\be L = f(\dot x + a)^2 + {x^2\over 2} + {\dot\phi^2\over 2} - V + \lambda(\dot a-x).\ee
Canonical momenta are,
\be p_x = 2f(\dot x + a),\;\;p_a = \lambda,\;\;p_{\phi} = \dot \phi,\;\;p_{\lambda} = 0.\ee
The constrained Hamiltonian then reads

\be H_c= {p_x^2\over 4f} - ap_x + {p_{\phi}^2\over 2}+ \dot \lambda p_{\lambda} - {x^2\over 2} + V + \lambda x.\ee
Clearly we require two primary constraints involving Lagrange multiplier or its conjugate viz,

\be \theta_1 = p_a - \lambda,\;\;\theta_2 = p_{\lambda}.\ee
The primary Hamiltonian therefore reads

\be H_{p1} = {p_x^2\over 4f} - ap_x + {p_{\phi}^2\over 2} - {x^2\over 2} + \lambda x + u_1(p_a - \lambda) + u_2p_{\lambda}.\ee
In the above $u_1$ and $u_2$ are Lagrange multipliers and the Poisson bracket $\{a, p_a\} = \{x, p_x\} = \{\lambda, p_{\lambda}\}= 1$, hold. Now
constraint should remain preserved in time. Therefore

\be\begin{split}& \dot\theta_1=\{\theta_1, H_{p1}\} = -u_2 + p_x + \sum_{i=1}^2\theta_i\{\theta_1,u_i\} =0\\&
\dot\theta_2=\{\theta_2, H_{p1}\} = (u_1-x) p_{\lambda} + \sum_{i=1}^2\theta_i\{\theta_2,u_i\}=0\end{split}\ee
Since, Constraints must vanish weakly in the sense of Dirac, therefore $u_1 = x$ and $u_2 = p_x$. The modified primary Hamiltonian therefore reads

\be H_{p2} = x p_a - a p_x + {p_x^2\over 4 f} +{1\over 2}p_{\phi}^2 - {1\over 2} x^2 + V + p_x p_{\lambda}.\ee
Now since $\dot\theta_2 = \{\theta_2, H_{p2}\} = 0$, so $p_{\lambda} = 0$. Hence, the phase-space Hamiltonian finally takes the form,

\be  \label{d} H_D = x p_a - a p_x + {p_x^2\over 4 f} +{1\over 2}p_{\phi}^2 - {1\over 2} x^2 + V.\ee
The same phase-space Hamiltonian is produced following Horowitz formalism also. However, the so called legitimate canonical formulation lead to
\be  \label{l} H_L = x p_a +{p_x^2\over 4 f}+{1\over 2}p_{\phi}^2 +2 f' a x p_{\phi} - {1\over 2}[(1-2f'a)^2-4f]x^2 -fa^2+ V\ee
While, Hamiltonian \eqref{d} does not require, the fourth term in the above expression \eqref{l}, requires operator ordering, which is different for different forms of $f(\phi)$. Clearly, therefore, $H_D$ and $H_L$ are not related through canonical transformation. Such situation is encountered due to non-minimal coupling which appears in the case of any non-minimally coupled higher order theory of gravity.

\end{document}